# Toward a New Approach for Modeling Dependability of Data Warehouse System

Imane Hilal
RITM Lab., CED Engineering Sciences ENSEM
ESTC, Hassan II University
Casablanca, Morocco
imanehilal@gmail.com

Nadia Afifi
RITM Lab., Computer Engineering Department
ESTC, Hassan II University
Casablanca, Morocco
nafifi@est-uh2c.ac.ma

Reda Filali Hilali
RITM Lab., Computer Engineering Department
ESTC, Hassan II University
Casablanca, Morocco
filali@est-uh2c.ac.ma

Mohammed Ouzzif
RITM Lab., Computer Engineering Department
ESTC, Hassan II University
Casablanca, Morocco
ouzzif@est-uh2c.ac.ma

*Abstract*—**The sustainability of any Data Warehouse System (DWS) is closely correlated with user satisfaction. Therefore, analysts, designers and developers focused more on achieving all its functionality, without considering others kinds of requirement such as dependability's aspects. Moreover, these latter are often considered as properties of the system that will must be checked and corrected once the project is completed. The practice of "fix it later" can cause the obsolescence of the entire Data Warehouse System. Therefore, it requires the adoption of a methodology that will ensure the integration of aspects of dependability since the early stages of project DWS. In this paper, we first define the concepts related to dependability of DWS. Then we present our approach inspired from the MDA (Model Driven Architecture) approach to model dependability's aspects namely: availability, reliability, maintainability and security, taking into account their interaction.**

## I. INTRODUCTION

Data Warehouse Systems (DWS) are specially used by decision makers to analyze the status and the development of their organization *[1]*, based on a large amount of enterprise information integrated from heterogeneous Data Sources *[2]*. This information is organized following a multidimensional structure to make exploration and analysis easier for users. The DWS's architecture is composed from several layers: (i) Heterogeneous Data Sources (DS), (ii) ETL (Extraction/ Transformation/ Loading) process which extract and transform data from these DS, and load the information into DW ,(iii) Data Warehouse repository, (iv) Restitution Tools that analyze the data in OLAP way.

The final goals to implement the DWS are: (i) To evaluate complex queries without causing severe impact on the sources; (ii) To increase data availability and decrease response-time for OLAP queries; (iii) To provide correct historical trends for state-oriented data; (iv) To protect and secure the crucial business information; (v) To back up modifications and insure evolution and maintenance. Those goals involve the guarantee of non-functional requirements such as availability, reliability, security, confidentiality, integrity and maintainability which are encompassed in the dependability's attributes [3].

To meet those goals, we suggest an approach that spans over dependability's attributes. Our contribution presents, on the one hand, the advantage of considering these attributes from the early stages of the DW project. On the other hand, it provides models to the designers and developers in order to realize these functionalities in respect to the Model Driving Architecture's (MDA) principles.

The remainder of this paper is organized as follows: section 2 will present the related work on the development of dependable DWS; section 3 will give an overview of DW, dependability aspects and MDA; section 4 will introduce our approach through which we develop our models for DWS dependability. An example of implementation is shown in section 5. Finally, in section 6 we present our conclusion and future work.

## II. RELATED WORK

In DW's literature, dependability's attributes have been neglected or they have been presented as a second class type of requirements, and have also been considered as the other non-functional requirements. But the experiences show that capturing non-functional requirements without mapping them into the conceptual model may provoke loss of information [4]. Only few works have specifically addressed this issue. In particular, Pain &castro [5] proposed (DWARF) the Data Warehouse Requirements Definition approach that adapts a requirement engineering process for requirements definition and management of DW. Their approach demands particular attention to non-functional requirements that are captured through an extension of the NFR framework. The same authors have provided an exhaustive classification of non-functional requirements that must be addressed in the development of DWS [6].

A different proposal comes from V. Peralta, A.Illarze, R. Ruggia, [7] who have suggested modeling the non-functional requirements through guidelines that are not directly related with the conceptual model. Instead, non-functional requirements have been used during logical design, where most of the choices related to the performance and security have already been taken.

E. Soler, V. Stefanov, J. N. Mazón, J. Trujillo, E. Fernández-Medina, M. Piattini, [8] proposed a security requirement model for DW based on the MDA approach[9]. This approach was extended by Carlos Blanco [10] who has developed a secure DW, including security issues in each stage of the development process. This is the only approach that considers security as a non-functional requirement from the early stages of the DW development.

On the other hand, some aspects of dependability such as reliability, availability, security and maintainability have been widely discussed in the DWS[11]. However, their interaction and integration were rarely discussed [11, 12].

## III. DEPENDABILITY'S ATTRIBUTES OF DATA WAREHOUSE SYSTEM

### A. *Dependability's attributes*

The original definition of dependability for a computing system is the ability to deliver service that can legitimately be trusted [13]. This definition induces the need for justification of trust to avoid service failure and make it more acceptable.

Dependability may be viewed according to different, but complementary attributes, which are summarized in: (i) Availability: ability of the system to be operational at the time requested (ii) Reliability: continuity of service, (iii) Security: non-occurrence of environmental catastrophic consequences, (iv) Confidentiality: the prohibition of unauthorized disclosure of information, (v) Integrity: non-occurrence of improper alteration of information, (vi) Maintainability: the ability to handle repairs and updates [13]. Associating integrity and availability together, as well as confidentiality, lead to security.

### B. *Data Warehouse System*

DWS are known as a collection of decision support technologies, aiming at making decisions better and faster. ETL (Extraction, Transformation, Loading) retrieves data from various operational databases, cleans, transforms and loads it in DW repository [2]. The rules used to clean and transform the data and perform the transformation are part of the metadata management system. Users can extract data from the DW using, for instance, query tools. The components of a typical DWS are shown in "Fig.1".

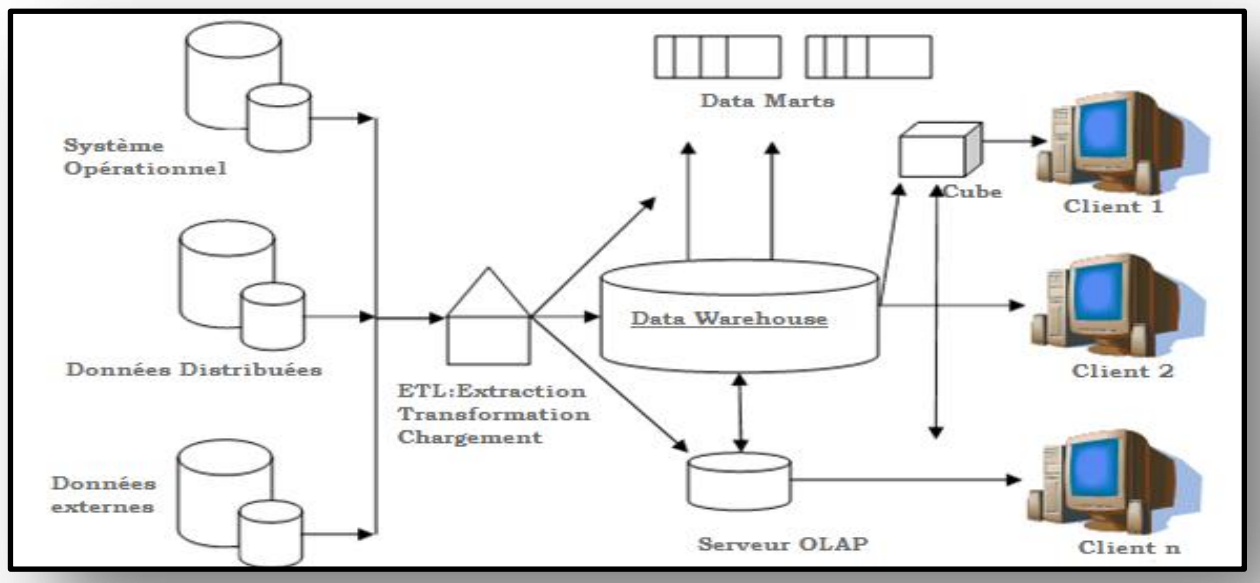

Figure1.Data Warehouse System's architecture

### C. *DWS's dependability*

In this section we will adapt each dependability attribute in the DWS context:

#### 1) *Availability*

Availability is the readiness of the system to deliver correct service whenever solicited [13]. The same definition is projectable on DWS. Thus, we can define the availability of the DWS as the latter's ability to provide extracted, cleaned, rebuilt information whenever needed, and which can be analyzable following different axes.

Literature provides many optimization techniques to deal with the demand of availability in DWS. Most of these techniques focus on DW, and are inherited from traditional relational database systems. Among these techniques [14] we find: materialized views, indexing methods, data partitioning [15], and parallel processing [16]. Regarding ETL, many techniques have been developed to improve its availability as it's mainly responsible for transforming and delivering data. Those techniques have proposed to optimize the ETL process, using the workflow concepts [17] and the concept of near-real-time data warehousing [18, 19].

#### 2) *Reliability*

The common Reliability's definition consists of the continuity of correct service [13]. Projecting this definition to DWS allows us to consider this attribute as the percentage of time the DWS is available for use. While considering aspects of maturity, fault tolerance and recoverability.

Considering the definitions of availability and reliability, both emphasize the avoidance of failures, and can be grouped together, and collectively defined as the avoidance or minimization of service outages.

#### 3) *Maintainability*

Maintainability is the ability to bear repairs and updates [13]. In reality, DWS are interacting with a dynamic environment due to changing business rules or operational objectives on one hand, and increasing requirements and changing in its data sources on the other hand. Thus, their maintenance is very important to insure their ability of adaption to their environment's evolutions.

Based on the previous work, we have to distinguish between two concepts:

- Maintainability: which is the ability of an element to be maintained; this ability stems from all design features that enhance the ability of service.
- Maintenance: is a series of actions of appropriate character (content, timing, quality) to restore or keep an element in an operational state.

Based on the definitions above, the two concepts are not contradictory, since maintenance is the action of maintaining. More the system accepts maintenance's operation, the more it's maintainable. In our proposal, we consider that the two aspects have the same meaning.

#### *4) Security*

Security is a basic requirement for diverse kinds of applications. Since business information in the data warehouse is sensitive, the different aspects (availability, integrity and confidentiality) of security are particularly important. In this way, information security should be taken into consideration from the early stages of DWS's development life-cycle [9, 10].

To summarize, the integration of dependability aspects is a necessary to insure a dependable DWS. But the challenge in this context is that the attributes are often: (i) Heterogeneous; (ii) Ranked differently and subjectively by the stakeholders; (iii) Interrelated: in conflict or in harmony. In addition, the implementation of these attributes requires a lot of expertise and domain knowledge.

## IV. MDA : MODEL DRIVEN ARCHITECTURE

MDA [20] provides model-driven software development based on the separation of the specification of the system functionality and its implementation. It defines models at different abstraction levels, and transformations required to the passage of a level to another [21] "Fig. 2".

This approach propose business models (CIM: Computation Independent Model) based on system requirements. Conceptual models (PIM: Platform Independent Model), which do not include information about specific platforms and technologies. These latter are including in the logical models (PSM: Platform Specific Model). Developing information systems applying the MDA approach improves productivity, saving time and effort, and provides support for system evolution, integration, interoperability, portability, adaptability and reusability [22].

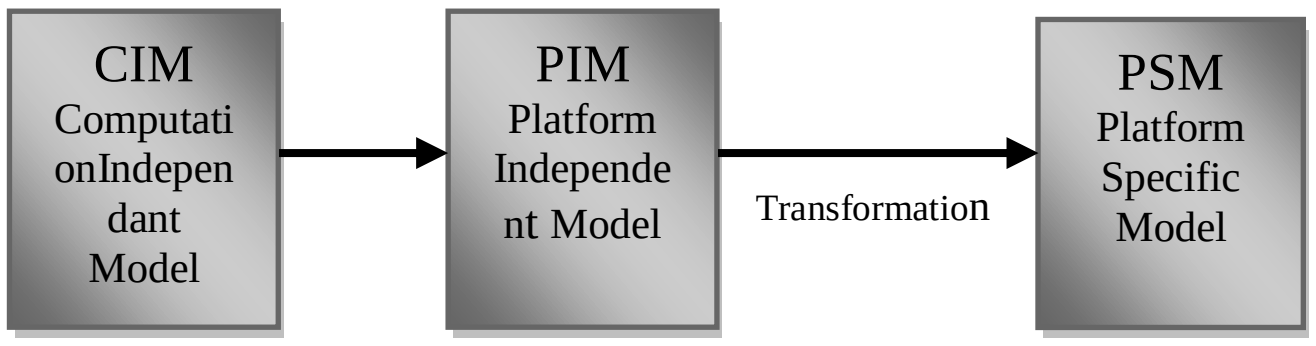

Figure 2. Generic MDA Models

The requirements of the future system are described in the CIM, which is refined into the PIM. The PSM is the result of the PIM transformation. The process refines the PIM based on the specification described in the Platform Description Module (PDM) defining how to use a specific platform.

The main advantages [21] of the MDA framework are: (i) Results are automatically generated which is expected to improve productivity, development duration, and cost; (ii) The developer pay more attention to CIM and PIM instead of focus on logical and technical details; (iii) PIM is portable to different target platforms; (iv) Once a transformation has been developed, it can be reused whenever needed; (v) Changes have to be done in the PIM only if the target platform has changed; (vi) New requirements in the CIM are passed from PIM to PSM immediately and changes are reflected automatically [14], [15], [16].

## V. OUR APPROACH

Our proposal aims to provide a dependable DWS process design, taking into account, since the early stages of modeling, dependability's constraints. As these latter vary according to the field of use of DWS, we suggest a generic approach aligned to the standard MDA, allowing their integration in a refined and incremental way. Of course it based on different levels of abstraction to ensure traceability, and check their compatibility while considering their interaction. "Fig. 3" explicitly represents our approach.

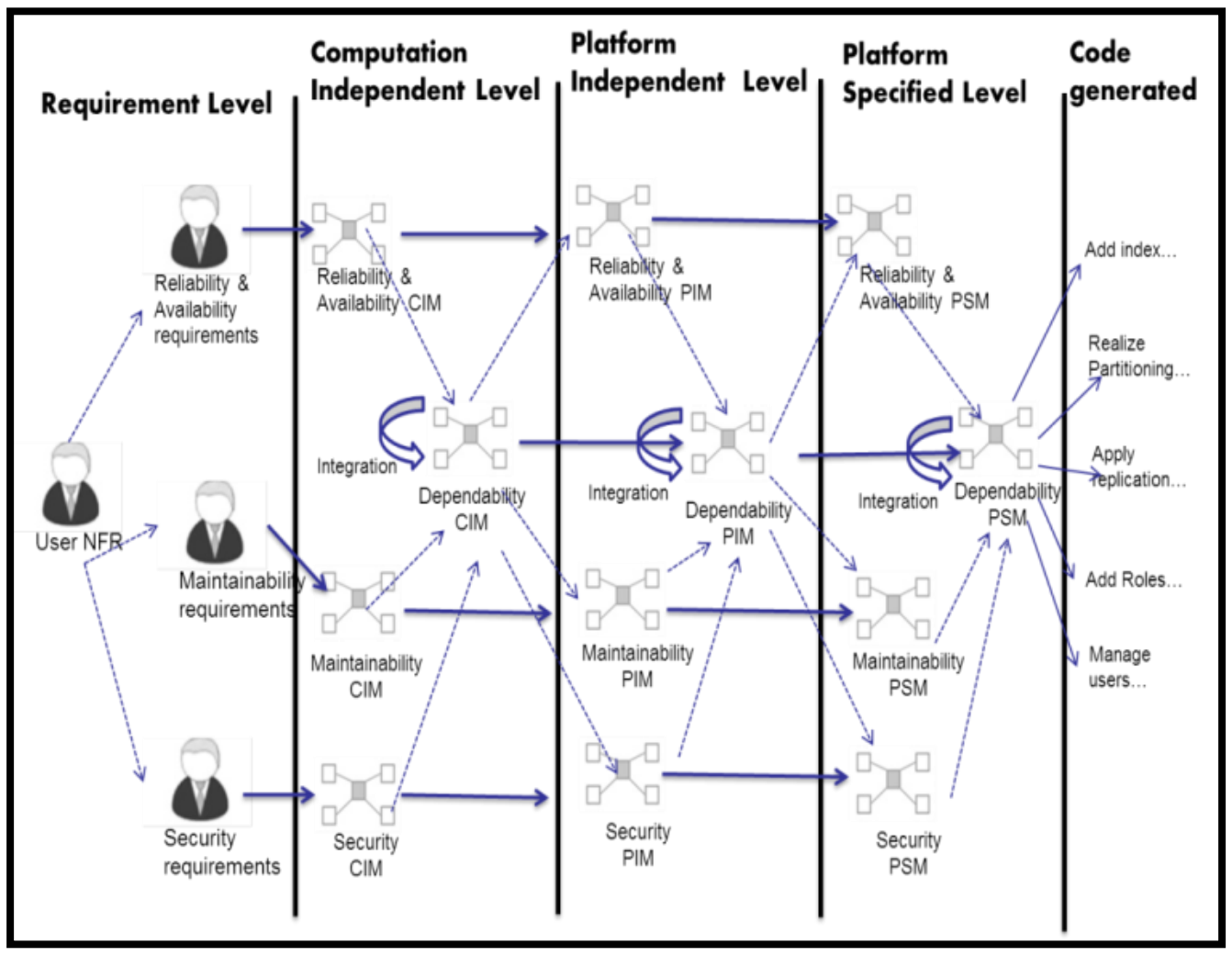

Figure 3. Our proposed approach

Our approach covers the entire DWS project. However, the DWS is itself a heterogeneous system, because of the platforms from which it is composed (DS, ETL, DW, analysis tools), we adapted the "divide to conquer" strategy. On one hand, it allows to deal with each layer preserving its own characteristics, and secondly integrate them with the surrounding layers. We have thus split the standard architecture of DWS into four homogeneous layers: the DS, ETL, DW and analysis tools. Then we conducted a qualitative study to list all parameters influencing the aspects of dependability. We have integrated them through building a structured approach respecting the different types of models proposed by the MDA approach.

### A. Requirement Level

This section is devoted to the qualitative study of all dependability's attributes in the four layers of the DWS. We isolated each of its layers. Based on the fact that the availability and reliability are very close (see section 3). They will therefore be associated in the following of our work. We can list the parameters that influence dependability's attributes considering that their implementation depends on the type of their attribution.

#### 1) Data Source (DS):

Availability and Reliability <Structuration, Volumetric, Optimization, Data Quality, Support, Replication> as:

- •Structuration <structured, mi structured, not structured>
- •Volumetric <transaction frequency, recording number >
- •Optimization <index, fragmentation, view, parallelism >
- •Data Quality <redundant, missing, incomplete, mistaken>
- •Support <hard disc, communication>

Maintainability: <Documentation, Structuration, Optimization, Support> as:

- •Documentation <schema, metadata>
- •Structuration <structured, mi structured, not structured>
- •Optimization <index, fragmentation, view, parallelism >
- •Support <hard disc, communication>

Security: <User Management, Data, Support> as:

- •User Management< authentification, permission, quotas, audit, hierarchy>
- •Data <integrity, authenticity, completeness, locking>
- •Support <firewall, encryption, replication>

#### 2) Extraction, Transformation, Loading (ETL)

Availability and Reliability <Data Quality, Periodicity, Cleaning, Complexity, Support> as

- •Data Quality <redundant, missing, incomplete, mistaken>
- •Periodicity <batch, real time, cycle refreshment>
- •Cleaning <consolidation, formatting, restructuration, normalization, doubloon elimination, aberrant values elimination, missing values >
- •Complexity < data size, calculation rules, DS number>
- •Support <RAM, temporary storage, communication>

Maintainability: <Metadata, Modularity, Complexity, Restoration> as:

•Metadata <source, destination, cleaning operations, management rules, constraints>
•Modularity <interdependence transformations, subsystems decomposition >
•Complexity <calculation rules, DS number>
•Restoration <recovery management, exception management>

Security: <Intrusion Management, Audit, Restoration> as:
•Intrusion Management <firewall, encryption, codification, access control >
•Audit <data, transformation>
•Restoration <recovery management, exception management>

#### 3) Data Warehouse (DW) Repository

Availability and Reliability <Schema, Volumetric, Complexity, Replication, Optimization, Support, loading Frequency> as:
•Schema <star, snowflake, constellation>
•Volumetric <recording number, access frequency, granularity>
•Complexity < fact number, dimension number, hierarchy number>
•Optimization <index, fragmentation, view, parallelism>
•Support <hard disc, communication>

Maintainability <Documentation, Architecture, Complexity, Optimization, Support> as:
•Documentation <metadata, schema, dimension>
•Architecture <centralized, distributed, consolidation Data Mart>
•Complexity <dependence dimensions, granularity>
•Optimization <index, fragmentation, view, parallelism>
•Support <hard disc, warm maintenance, cold maintenance >

Security <User Management, Data Management, Support> as:
•User Management <authentification, permission, quotas, audit, hierarchy>
•Data Management <integrity, authenticity, completeness, locking>
•Support <Firewall, encryption, replication>

#### 4) Restitution tools

Availability and Reliability <Type, Volumetric, Complexity, Support> as:
•Type < Ad hoc query, parameterizable query, dashboard, cube OLAP, Data Mining>
•Volumetric <data size>
•Complexity <axis number, transformation, granularity>
•Support <RAM, Communication>

Maintainability <Type, Documentation, Modularity, Complexity> as:

- •Type < Ad hoc query, parameterizable query, dashboard, cube OLAP, Data Mining>
- •Documentation <metadata, applications, indicators, calculation rules>
- •Modularity <interdependence, parallelism>
- •Complexity <Axis number, transformation, granularity>

Security: < User Management, Data Management, Support> as:

- •User Management<authentification, permission, quotas, audit, hierarchy>
- •Data Management <integrity, validity, privacy>
- •Support <firewall, encryption>

## *B. Computation Independent Level*

At this stage, we will implement the Computation Independent Model (CIM) using the NFR (Non Functional Requirements) Framework approach. This model is based on the parameters from the previous level. The NFR Framework proposed by Chung [23], is based on the concept of "Softgoal" to represent information about the different qualities of a system, and their interdependencies [24].

The development of interactive SIG (Softgoal Interdependency Graphs) allows representing the dependability's attributes and their integration. Then, it can examine the interactions and identify potential conflicts, to better suggest a compromise in the early stages of the DWS project. The following diagrams “Fig. 4, 5, 6” clarify the CIM availability & reliability, security and maintainability of DWS.

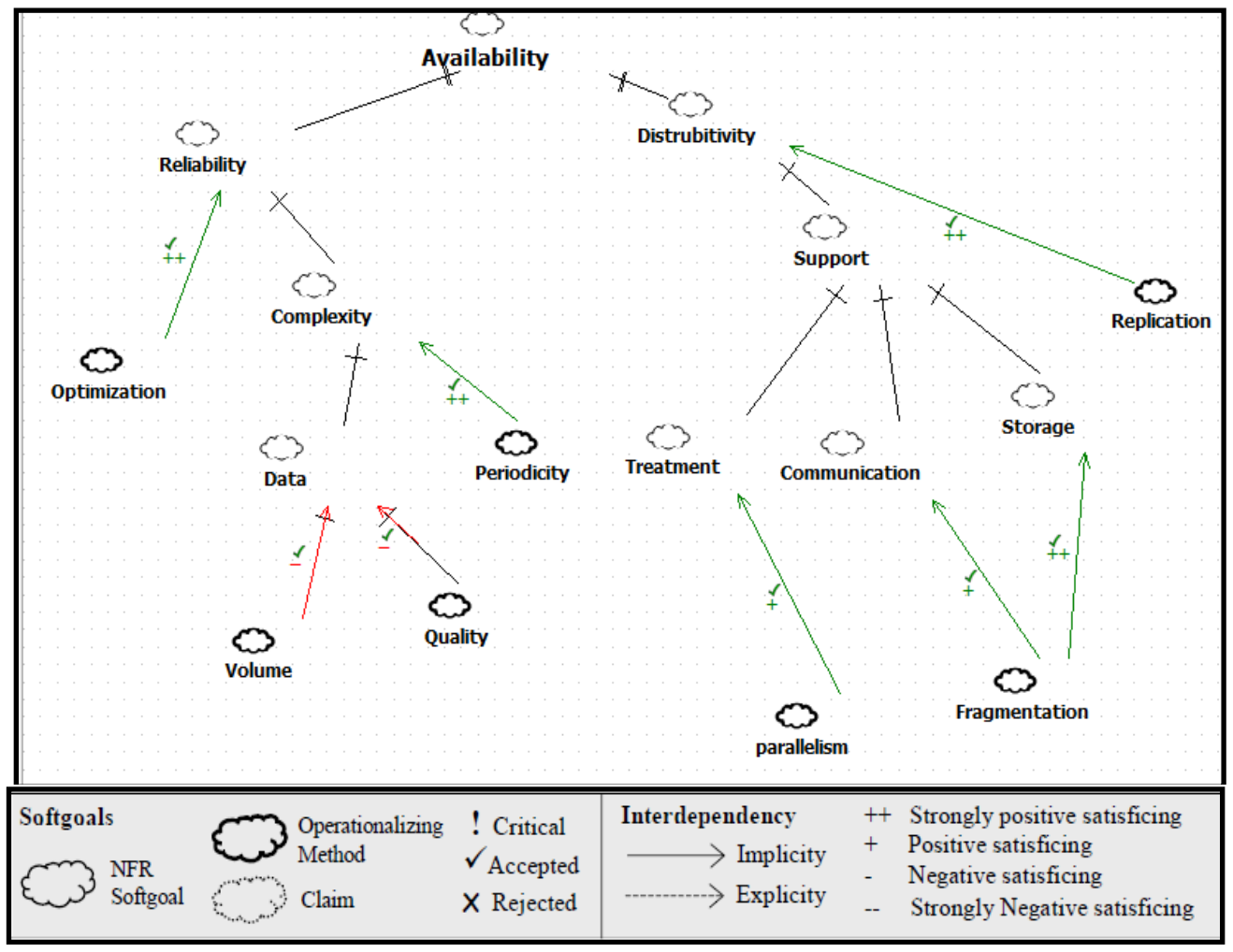

Figure 4. Availability & Reliability CIM

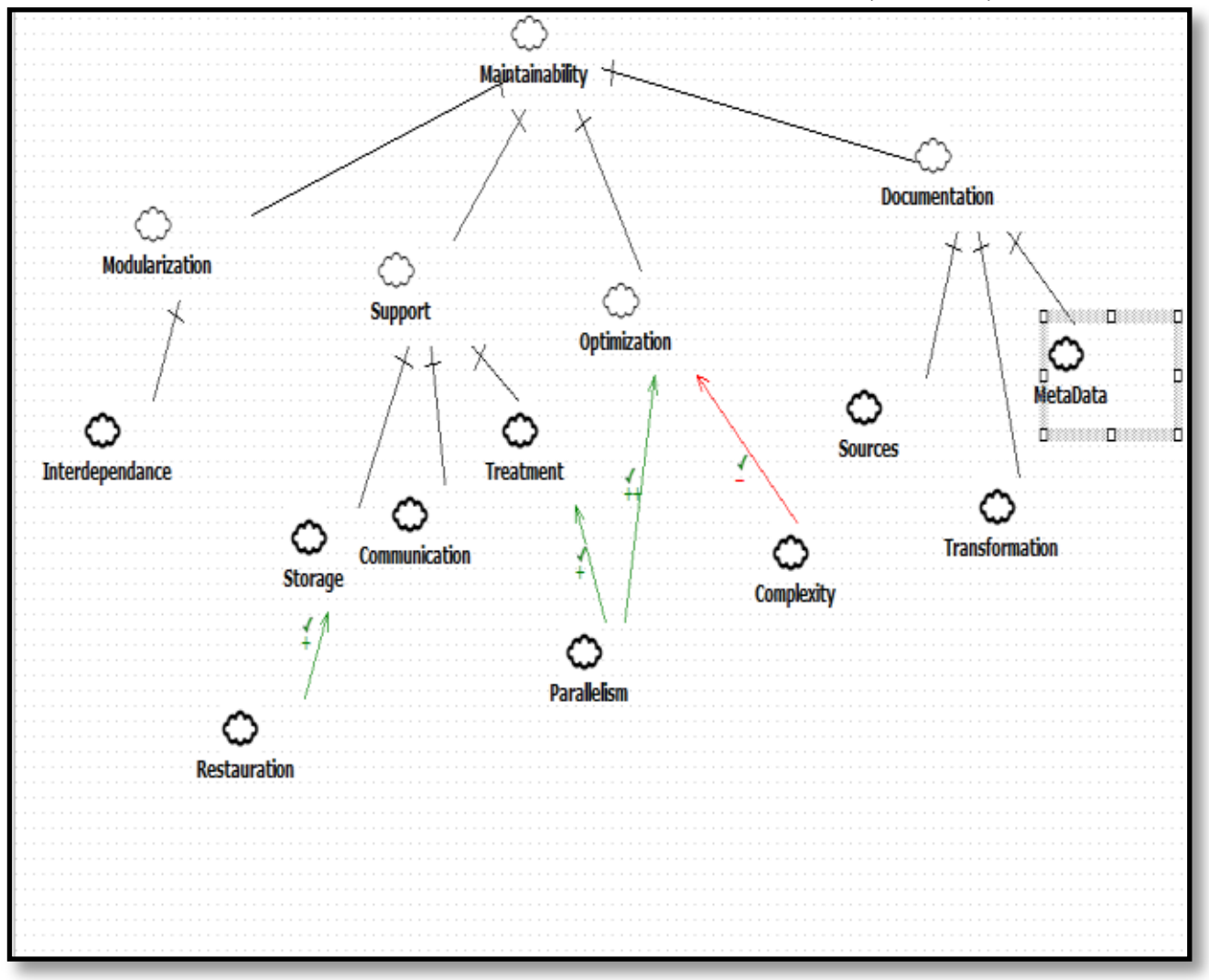

Figure 5. Maintainability CIM

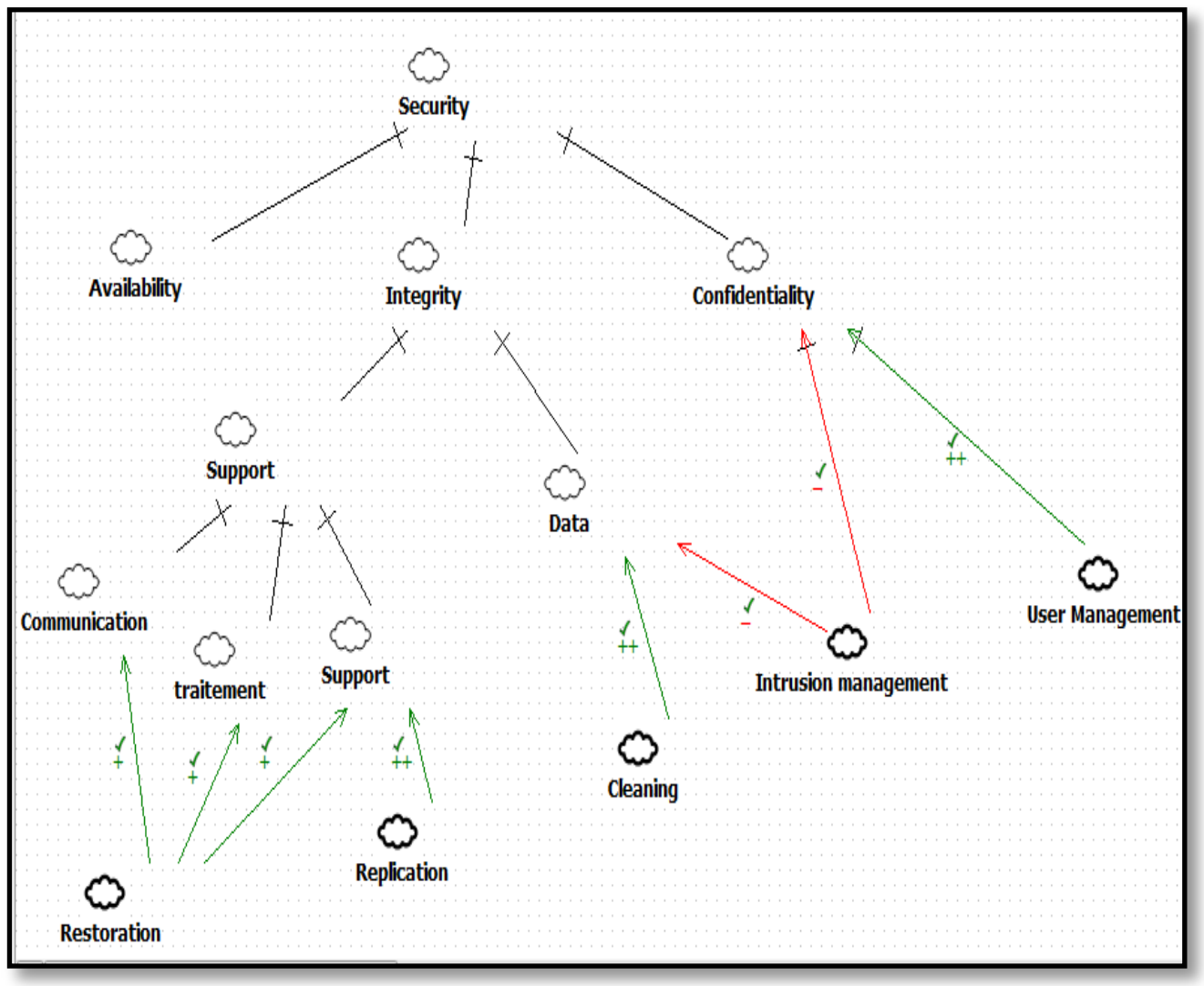

Figure 6. Security CIM

“Fig.7” shows an example of integration of the above models (“Fig. 4, 5, 6”). This integration will allow the designer to view all the dependability’s constraints in a model. Therefore, the appropriate implementation’s decisions of functional modules will be made taking into account non-functional requirements relating to dependability’s one.

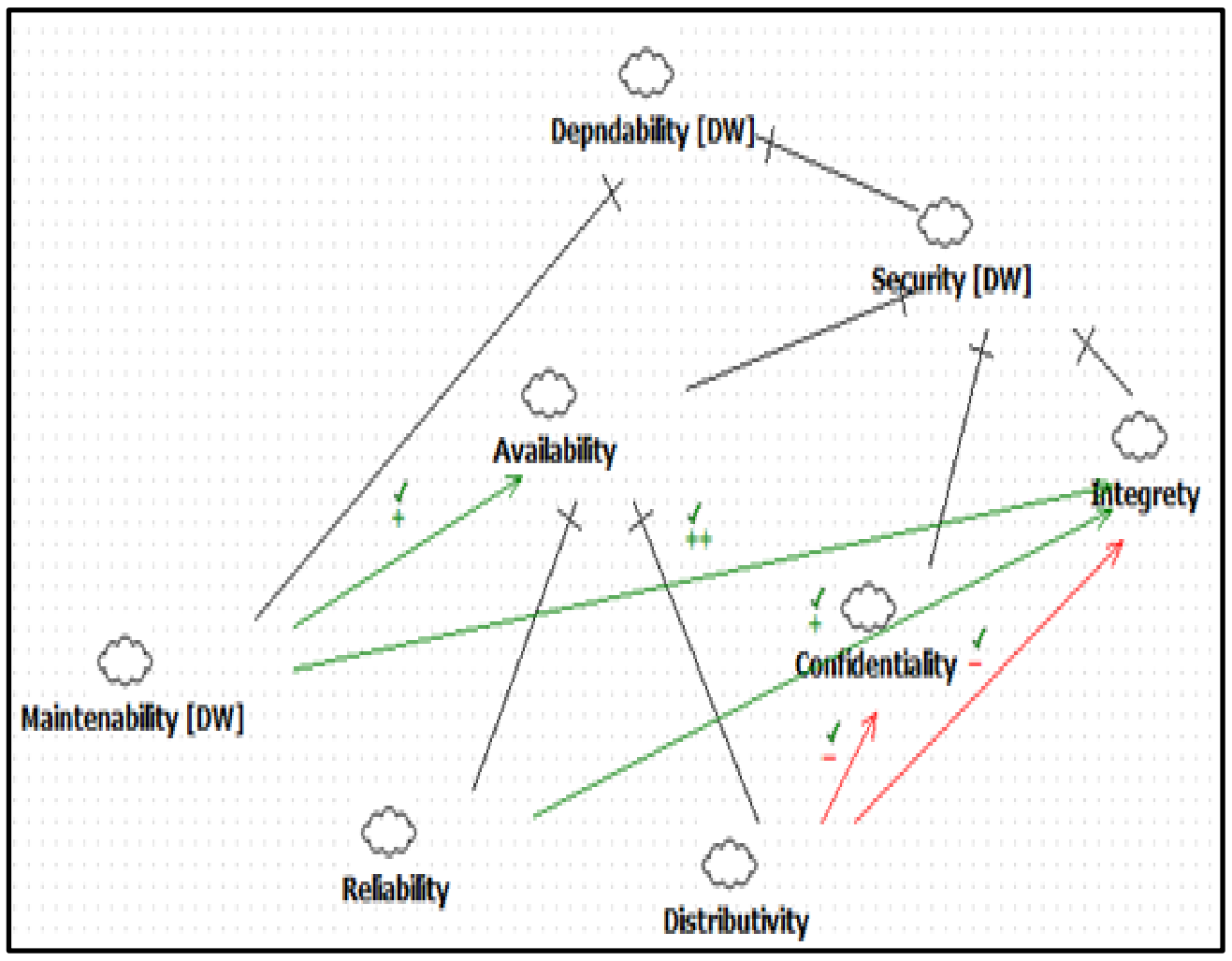

Figure7. Example of CIM integration

### C. *Platform Independent Level*

#### 1) *The transition from CIM to PIM*

To ensure traceability of dependability's aspects since the expression of requirements to the implementation, we was based on the work of S. Supakkul, L.Chung [25]. The authors proposed an integrated modeling language by extending UML with the NFR framework using the UML profile. They defined a MetaModel to represent the concepts in the NFR Framework. They alsoidentified the extension points for associating the two notations.

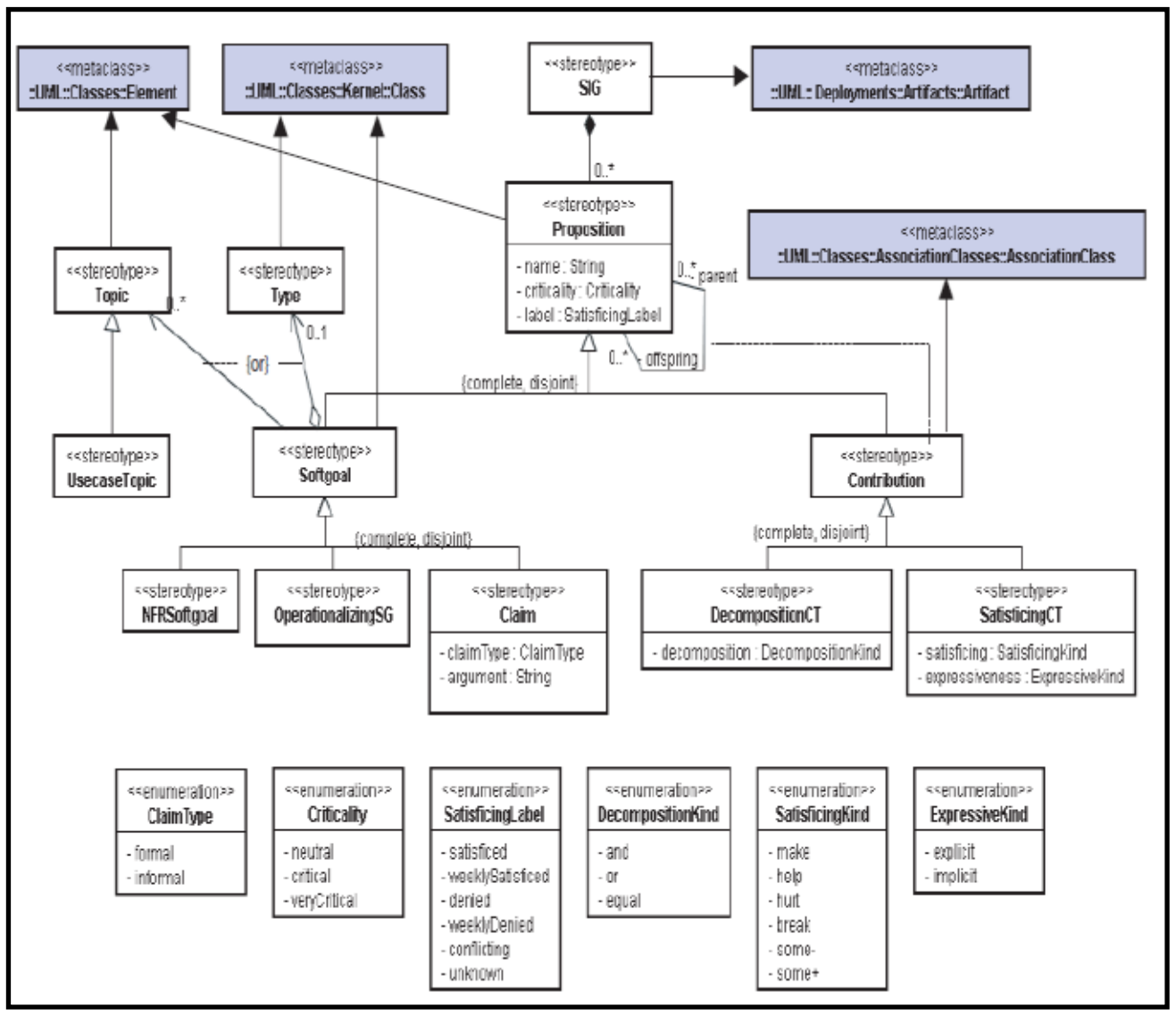

Figure 8. MetaModel of NFR Framework & relationship with UML metadata from [25]

"Fig. 8" explains the relationship between the Meta model of NFR Framework and the UML MetaModel. Therefore, we justified the existence ofthe transformation from our CIM (using NFR Framework) to PIM (using UML profile) by exploiting this relationship [26].

#### 2) *Development of PIM*

In this part we relied on an existing UML profile [27]: UML profile for Quality of Service "QoS" to represent the properties of quality of service that include dependability. "Fig. 9, 10, 11"represent the PIM of availability & reliability, maintainability and security according the profile already cited "Fig.12" shows an example of their integration.

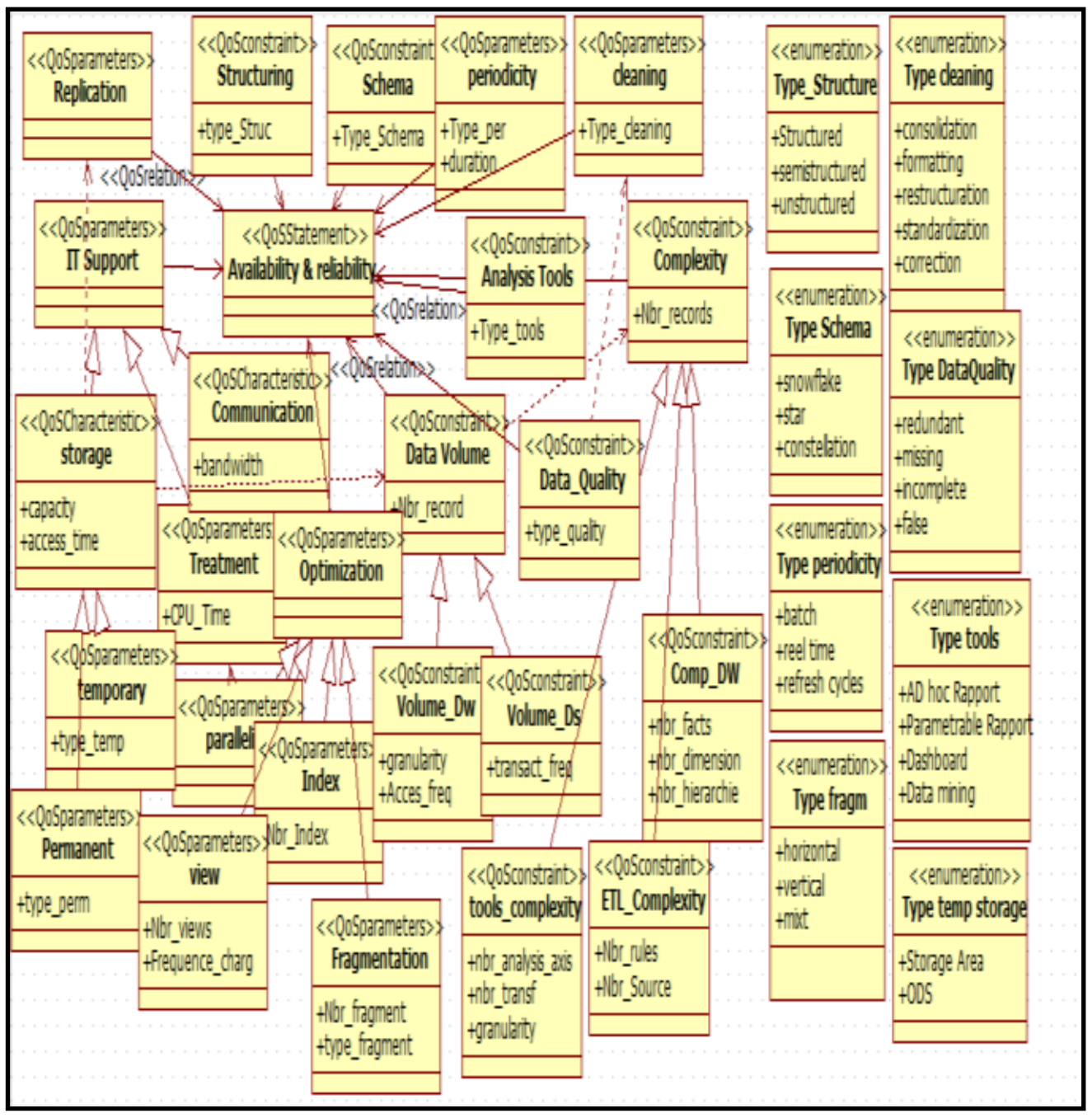

Figure 9. PIM's Availability & reliability

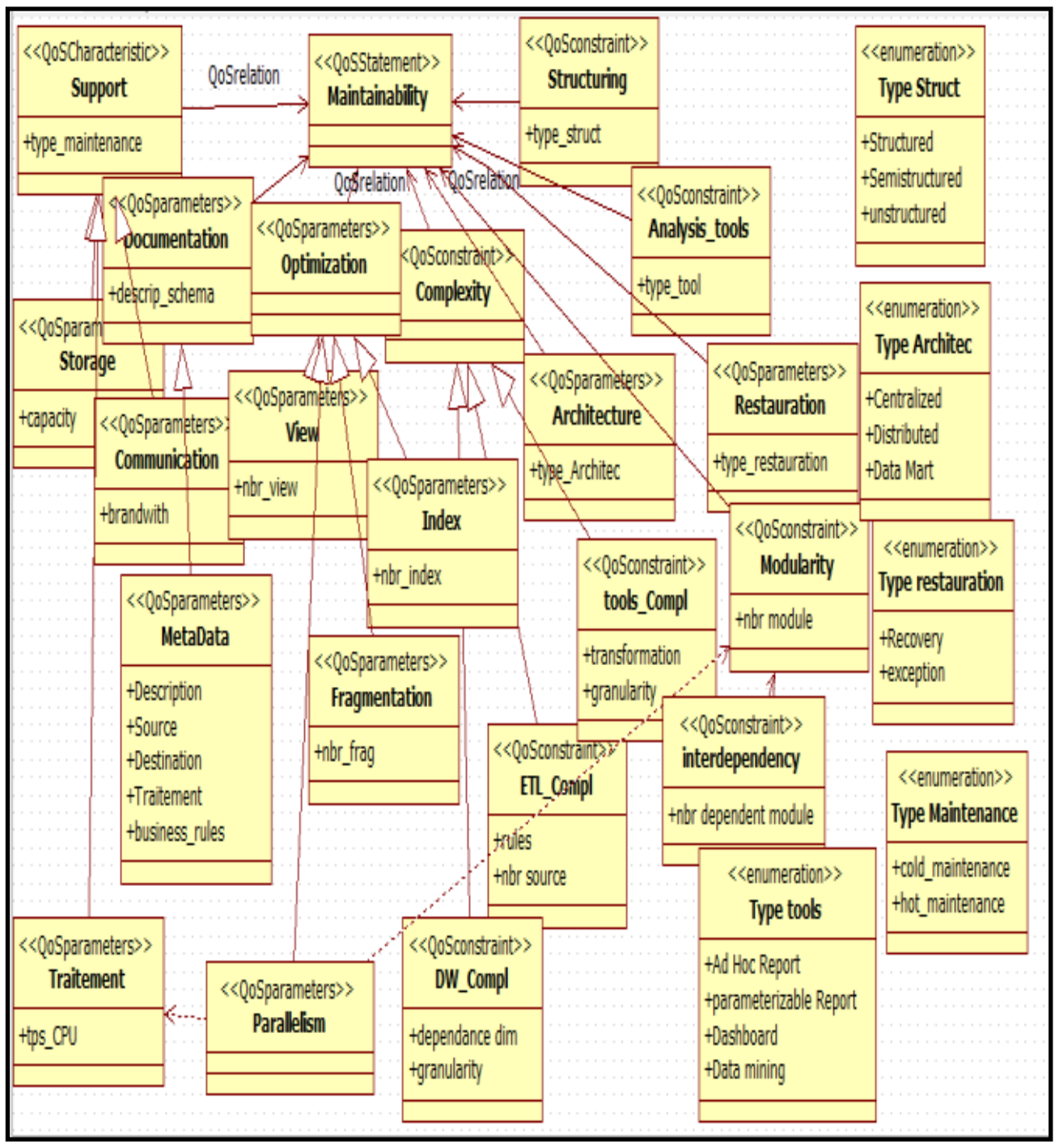

Figure 10. Maintainability PIM

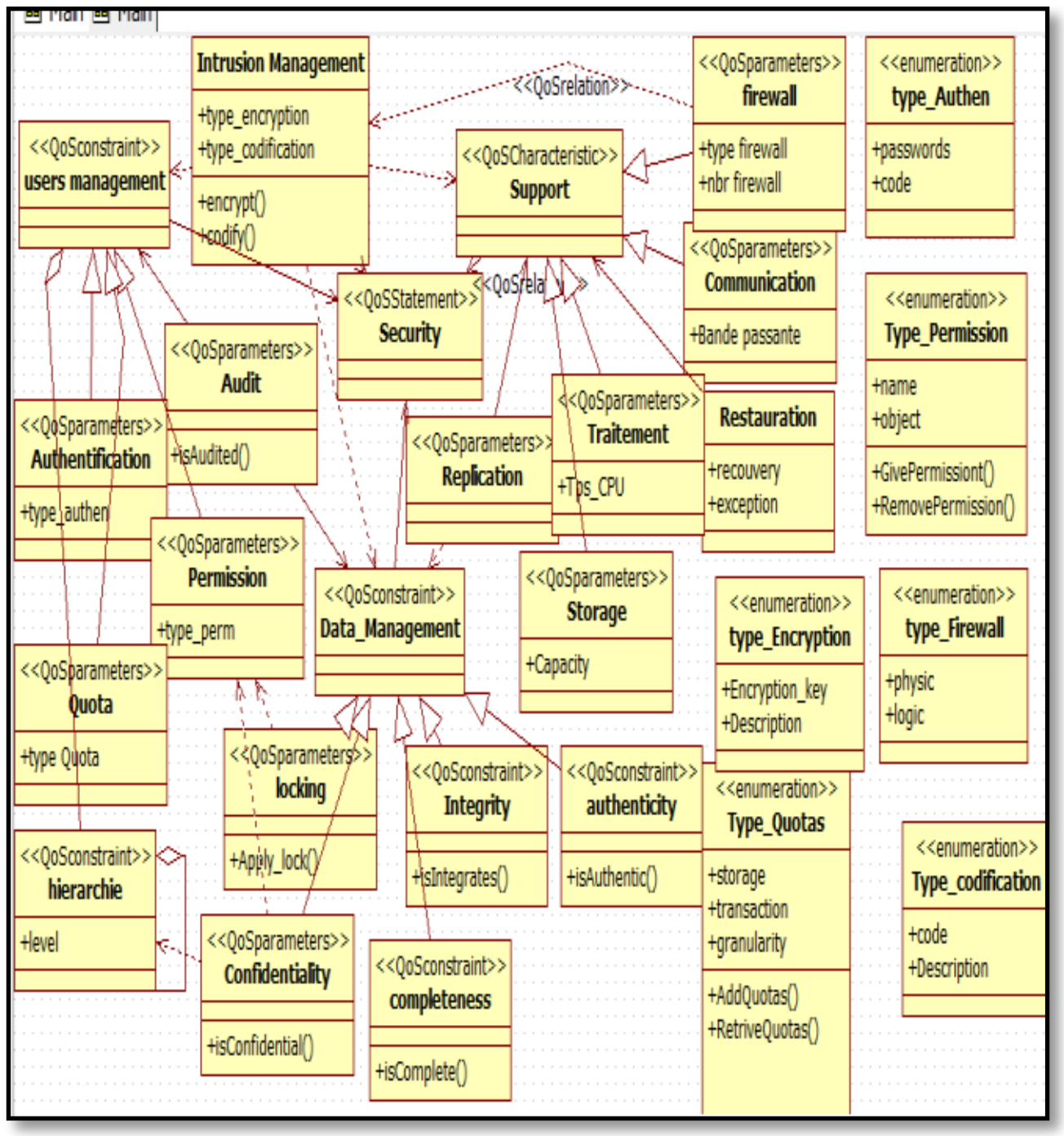

Figure 11. Security PIM

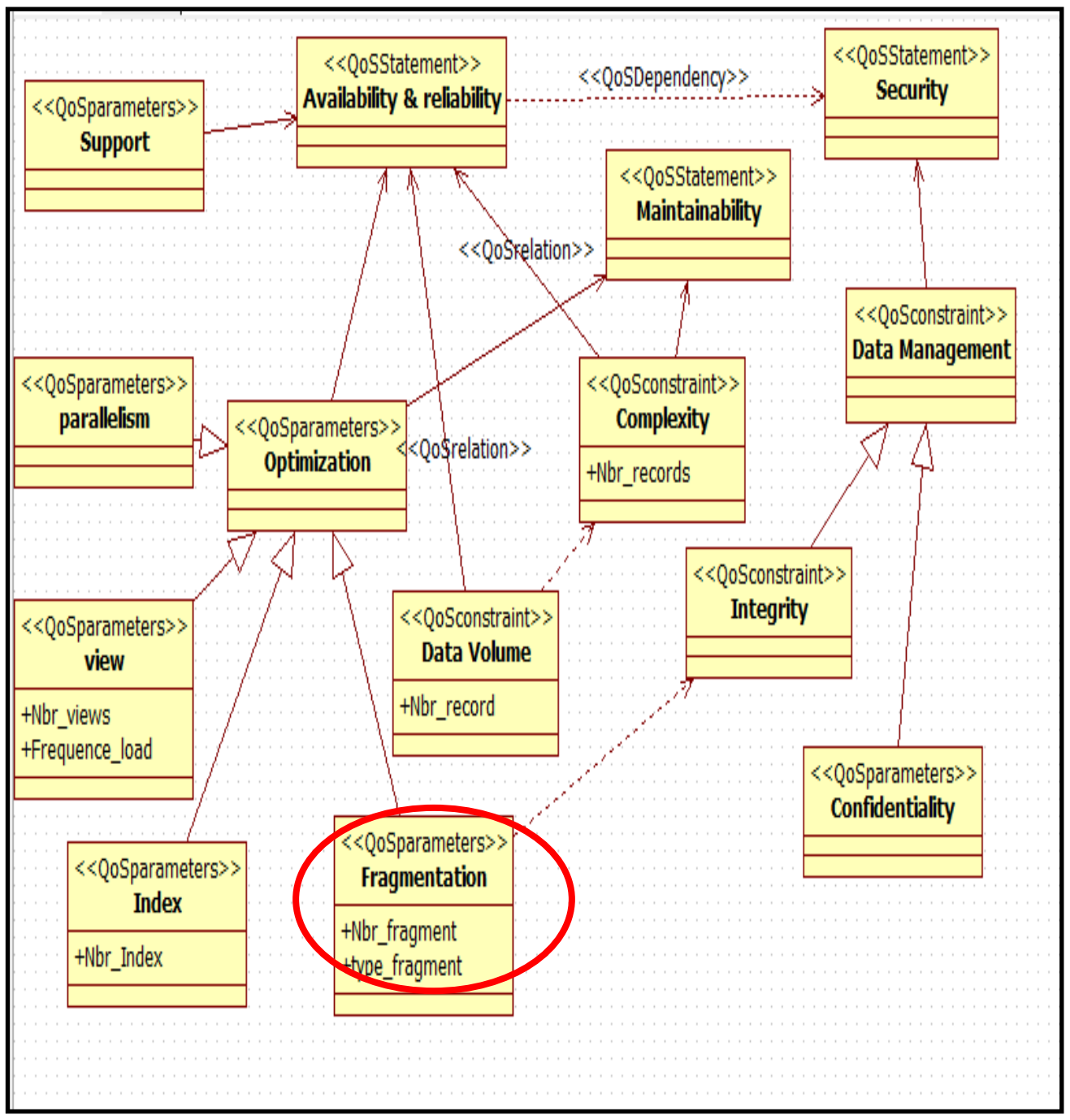

Figure12. PIM which integrate the dependability's attributes

## *D. Specified Level Platform*

At this level, developers need to combine information on the PDM platform (Platform Description Model) with PIM to generate the PSM. Given the heterogeneity of platforms which composed DWS architecture, we suppose that specific PSM must be developed for each of dependability attributes.

However the DWS is a system composed of several operatively different layers. Thus, for each layer of the DWS and every aspect of the dependability, a PSM must be establishedto keep the specifications of each layer. To keep our approach valid for all platforms, we will do not specify the PSM because they usea specific technology.

Once the PSM are refined, the generation code is automatic. In addition, the integration of dependability's attributes with the functional part is guaranteed through the aspect oriented programming. In the next section we present a case study to demonstrate the validity of our approach.

## VI. EXAMPLE OF IMPLEMENTATION

In our case study we take example of implementation of some of aspects of the Business Intelligence dependability, where users require as priority needs, data availability and reliability. However, it supposed that data are often maintained. Thus to ensure a dependable DWS we must take into account the interaction between the three aspects.

Based on CIM and PIM proposed by our approach, we were able to select the common parameters to the various attributes that are availability, reliability and maintainability, and on which we can act. These parameters are: replication, support and optimization (index, views, fragmentation, parallelism).

Support and replication technics can be directly applied on the media platform of DWS. We chose to implement fragmentation in the DW repository. The tests were performed by a workload query running on the APB-1 benchmark [28] under Oracle 10g. We applied the horizontal fragmentation of 10 queries Q1…Q10 of thebenchmark. The response time comparison of these queries before and after fragmentation is illustrated by "Fig. 13".

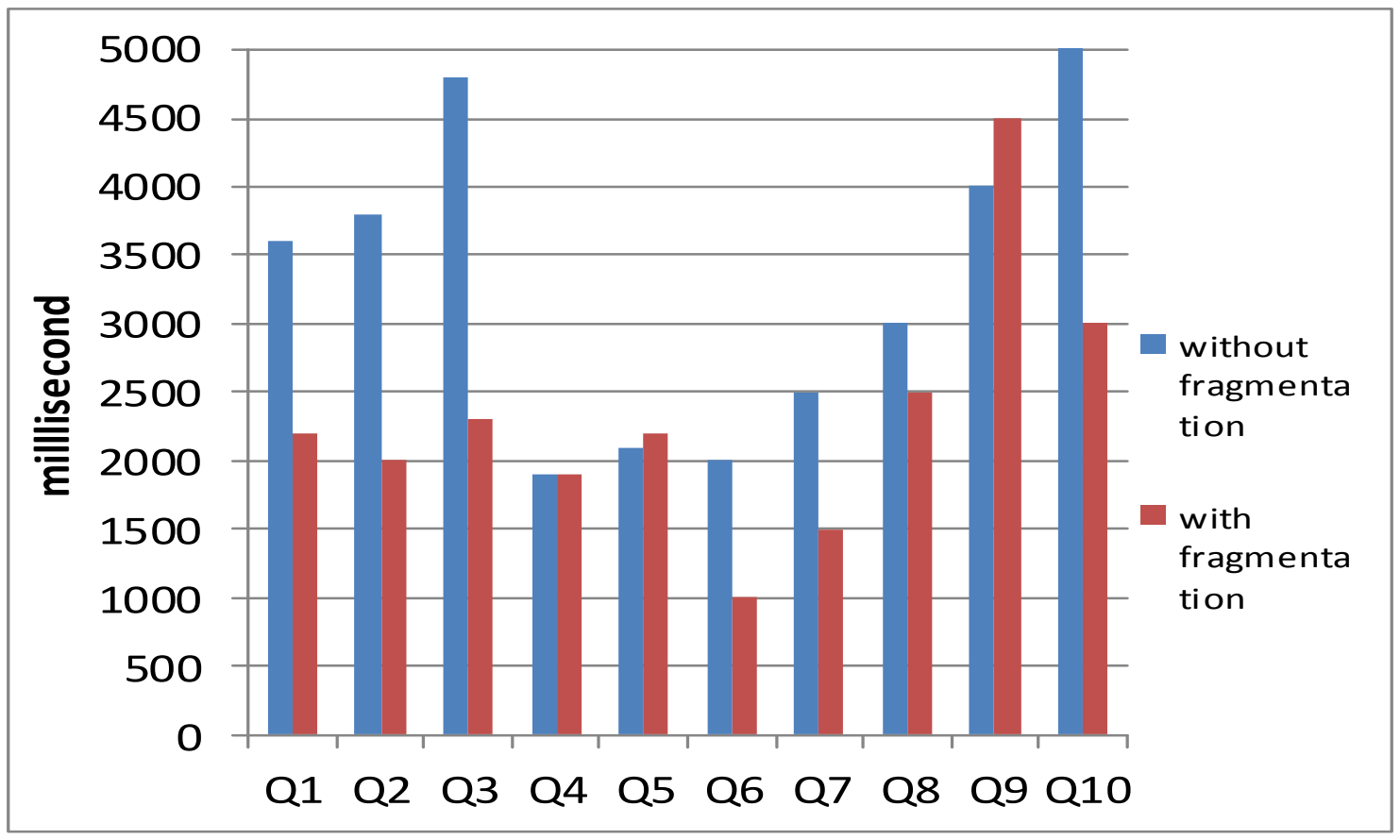

Figure 13. Graphical representation of results

In the following the summary of results:

- ✓ Fragmentation has improved the response time of the data warehouse which has a positive impact on the availability and reliability.

- ✓ Operations of maintainability and security can be slowed because of the complexity generated by the different fragments.
- ✓ We can adjust the number of fragments in order to find a compromise between availability, reliability and maintainability, security taking into account the levels tolerated by users.

## VII. CONCLUSION AND PERSPECTIVE

Our approach offers models at different levels of abstraction to deal with attributes of dependability. The main difficulty was to select measurable parameters common to the various aspects of the dependability. This problem has been overcome by our proposal, which integrates the different attributes taking into account their interaction from the early phases of modeling. To realize this, we used the models offered by the MDA.

As perspective, we can propose the integration of aspect-oriented programming approach in order to concretize the implementation of our proposal. We can also suggest the definition of an UML profile specific to dependability in DWS's context.